%% file: advanced.tex
\documentclass[twocolumn,letterpaper,10pt]{article}
\usepackage{iasted}
\usepackage{times}
\usepackage[dvips]{graphicx}
\usepackage{color,amsmath,amsfonts,amssymb,latexsym}
\definecolor{LightGray}{gray}{0.90}
\definecolor{MidGray}{gray}{0.80}
\definecolor{DarkGray}{gray}{0.25}

\definecolor{DeepRed}{rgb}{1.0,0.1,0.25}
\definecolor{DeepBlue}{rgb}{0.5,0.0,1.0}

\definecolor{DarkRed}{rgb}{0.65,0.15,0.0}
\definecolor{DarkGreen}{rgb}{0.15,0.65,0.0}

\definecolor{LightRed}{rgb}{1.0,0.40,0.40}
\definecolor{LightGreen}{rgb}{0.5,1.0,0.5}

\definecolor{LightBlue}{rgb}{0.5,0.5,1.0}
\definecolor{NavyBlue}{rgb}{0.1,0.0,0.9}
\definecolor{LightCyan}{rgb}{0.5,1.0,1.0}

\definecolor{LemonChiffon}{rgb}{1.,0.98,0.8}
\definecolor{LightYellow}{rgb}{1.0,1.0,0.35}
\definecolor{LightOrange}{rgb}{1.0,0.70,0.35}
\definecolor{MidYellow}{rgb}{0.8,0.8,0.0}
\definecolor{MidOrange}{rgb}{0.75,0.4875,0.0}
\definecolor{orange}{rgb}{1.0,0.65,0.0}
\long\def\symbolfootnote[#1]#2{\begingroup%
\def\thefootnote{\fnsymbol{footnote}}\footnotetext[#1]{#2}\endgroup} 
\begin{document}

\date{\today}

{\title{A neural model of the locust visual system for detection of object approaches with real-world scenes}}

\author{
Matthias S. Keil\footnotemark $\ \ $,$\ $ Elisenda Roca-Moreno, {\'A}ngel Rodriguez-V{\'a}zquez\\
\vspace*{-0.2cm}\\
{\small First published in:}\\
\vspace*{-0.2cm}\\
Proceedings of the Fourth IASTED International Conference\\
Visualization, Imaging, and Image Processing\\
September 6-8, 2004, Marbella, Spain
}
\maketitle
\symbolfootnote[1]{Corresponding author. Present address: Unviversity of Barcelona, Faculty of Psychology, Barcelona, Spain. E-mail: matskeil[AT]ub.edu}

\thispagestyle{empty}
\input{./definitions.tex}

\def\Rind{{\it Rind et al.}}
\def\ETA{\eta(t) \propto \dot{\Theta}\,\exp(-\alpha\Theta)}
\def\etafun{$\eta$-function}
\def\CITE#1{ \cite{#1}}
\noindent
{\bf\normalsize ABSTRACT}\newline
{In the central nervous systems of animals like pigeons and locusts, neurons were identified
which signal objects approaching the animal on a direct collision course.  Unraveling the 
neural circuitry for collision avoidance, and identifying the underlying computational
principles, is promising for building vision-based neuromorphic architectures, which
in the near future could find applications in cars or planes.  At the present there
is no published model available for robust detection of approaching objects under
real-world conditions.  Here we present a computational architecture for signalling
impending collisions, based on known anatomical data of the locust \emph{lobula giant
movement detector} (LGMD) neuron.  Our model shows robust performance even in adverse
situations, such as with approaching low-contrast objects, or with highly textured and
moving backgrounds.  We furthermore discuss which components need to be added to our model
to convert it into a full-fledged real-world-environment collision detector.} \vspace{2ex}

\noindent
{\bf\normalsize KEY~WORDS}\newline
{Locust, LGMD, collision detection, lateral inhibition, diffusion, ON-OFF-pathways, neuronal dynamics, computer vision, image processing}
\section{Introduction}
\label{sect:intro}
It is essential to many animal species to recognize and avoid approaching predators.
Animals like locusts or pigeons trigger collision avoidance behavior by relying
exclusively on monocular information.  Approaching objects give rise to an
expanding image on the animal's retina, subtending a visual angle $\Theta(t)$.  
From $\Theta(t)$, both the expansion rate $\dot{\Theta}(t)$ and the angular
acceleration $\ddot{\Theta}(t)$ can readily be computed.  Object approaches
with constant speed result in an approximately exponential increase
in $\Theta(t)$ and $\dot{\Theta}(t)$.  Such information may
in principle be evaluated in the visual systems of pigeons and locusts, 
in order to trigger avoidance reactions.  Whereas there is evidence that in pigeons
three collision-sensitive variables are computed in parallel \CITE{SunFrost98},
the locust seems to compute only one \CITE{Rowell71a,Rowell71b,OSheaWilliams74}.
Specifically, it has been found that responses
of the \emph{lobula giant movement detector} (LGMD) neuron correlate with object approaches
\CITE{RindSimmons92a,RindSimmons92b}.  However, the type of ''algorithm'' implemented
by the LGMD is a matter of ongoing debate, and currently prevailing hypothesis are suggesting
contrasting points of view concerning the functional roles of \emph{feedforward inhibition} 
(FFI) and \emph{lateral inhibition} (LI), respectively \CITE{RindSimmons99,GabHatsKrapp99,RindSimmons99b}
(notice that both types of inhibition act to suppress LGMD responses).
With the first hypothesis it was suggested that FFI accounts for the
LGMD's selectivity for approaching over receding objects, and for suppressing
LGMD responses due to self-motion of the organism.  A corresponding mechanism
relies on the concurrent activation of a large number of local movement detectors
(MDs) \CITE{RowellEtAl77}.
On the other hand, LI -- in conjunction with excitation due to MDs -- was proposed to implement a \emph{critical race}
over the LGMD ''input cables'' (dendrites) \CITE{RindBramwell96}:  in the early phase of an object approach,
and for translating objects, the activation of MDs over the photoreceptor arrays occurs
linearly (or slower than linear) with time.  Such activation patterns are canceled by
laterally propagating inhibitory waves (these waves are triggered from
previously activated MDs).  The latter situation is different from the late phase of
an object approach, where MDs are activated in a nearly exponential fashion.
While the LI wave propagates with constant speed, the wavefront of MD activation
spreads with continuous acceleration and thus can escape inhibition. Since the
(uncanceled) MD activation directly excites the LGMD, it finally responds.\\  
The second hypothesis was derived from the observation that LGMD responses to 
approaching objects could be fitted against an \emph{\etafun}, where $\ETA$, and
$\alpha=\mathit{const.}$ \CITE{HatsopoulosEtAl95}. The \etafun\s reveals an 
activity peak for objects approaching at constant velocity. Consequently, if
the LGMD computed something like an \etafun, then it would need to implement
the multiplication operator.  Indeed there is now neurophysiological evidence 
that multiplication of the excitatory input $\dot{\Theta}$ with the (feedforward) 
inhibitory input $\exp(-\alpha\Theta)$ is performed by first logarithmically encoding
both of the latter terms, before they are added in the LGMD neuron\CITE{GabKraKocLau02}.
It seems that the logarithmic encoding is subsequently undone in the
''output cable'' (axon) of the LGMD by voltage-dependent sodium conductances.
\begin{figure*}[ht!]
	\begin{minipage}[b]{\textwidth}
		\begin{center}\scalebox{0.99}{\includegraphics{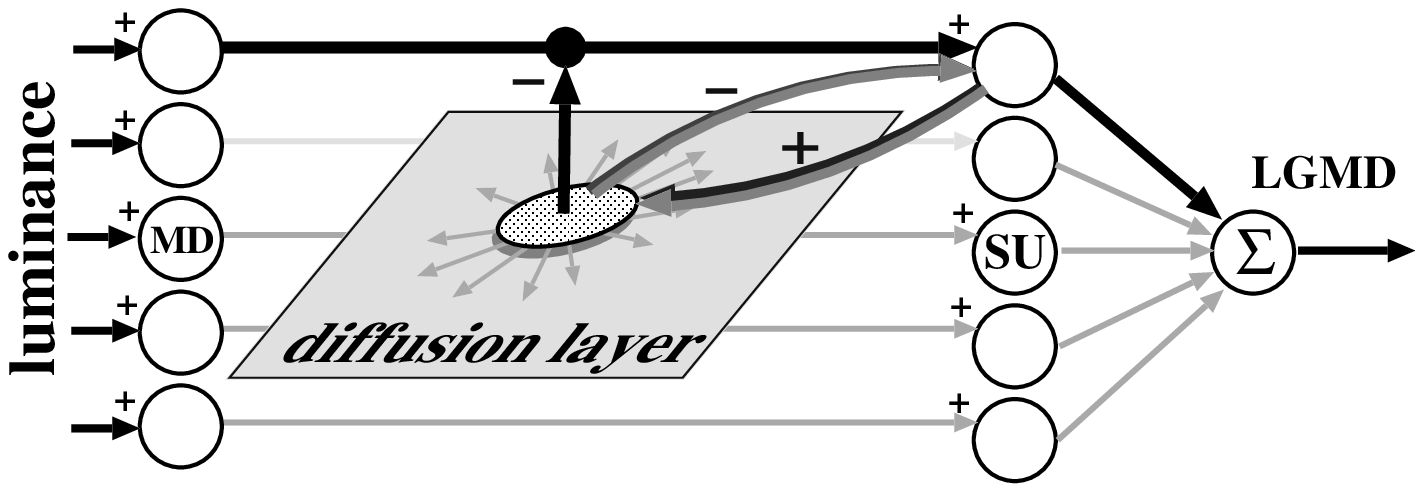}}\end{center}
	\end{minipage}
	\Caption[Sketch][Model diagram][{Movement information is extracted by movement detectors
	(MD, \eq[advancedphoto]) from a series of \emph{luminance} frames.  MD output feeds into
	summing units (SUs, \eq[summingUnits]).  SU output excites the \emph{diffusion layer}
	(\eq[advancedsyncytium]). Diffusion layer activity both decreases the input from MD into
	SU (vertical arrow, \eq[excinput]), and inhibits SUs (\eq[inhinput]). Finally, the \emph{LGMD} sums
	the activity of SUs.  LGMD activity represents the model's output.  Notice that there
	exist two independent pathways (ON and OFF) as the one sketched in the figure.  Plus
	signs at arrows indicate excitation, and minus signs indicate inhibition.  The horizontal
	arrow connecting the second (from above) MD with its corresponding SU has a lighter
	gray value only for reasons of improving the visualization.}]
\end{figure*}
\section{Model outline}
\label{sect:outline}
Our approach is based on the aforementioned ''critical race'', and represents
an improvement of a model previously presented in \CITE{MatsAngel03gc}.  It
processes information along two parallel streams, where one is sensitive for
luminance increments occurring from time $t-1$ to $t$ (ON), and the other one is
sensitive for luminance decrements (OFF).  The ideal output of the model
correlates with $\Theta(t)$ for object approaches (the nearer, the higher),
and is zero for all remaining movement patterns (such as generated by translating
objects, background movement, etc.).  The model is sketched in \fig[Sketch].
\subsection{ON- and OFF movement detectors}
The input into the model is provided by video sequences.  A frame of a sequence
at time $t$ corresponds to a luminance distribution $0\leq\Lumi[](t)\leq1$.
A movement detector neuron $p_{ij}$ at position $(i,j)$ is defined by
\begin{eqnarray}\label{advancedphoto}
\ddt[p_{ij}]  =  -\gLeak p_\mathit{ij} & + \Lumi[ij](t) (1-p_\mathit{ij})\\ 
                                       & - \Lumi[ij](t-1) (1+p_\mathit{ij})\nonumber
\end{eqnarray}
where $\gLeak=100$ is a passive decay, which is related to the degree of
low-pass filtering in time.  It also implements saturation of $p_{ij}$
with increasing input luminance.  The output (or activity) of an ON-movement detector
corresponds to positive values of $p_{ij}$, that is $\pos[{\rected[p]}]\equiv\rect[p]$,
where $\rect[\cdot]\equiv\max(0,\cdot)$ denotes half-wave rectification.  .
Negative values of $p_{ij}$ encode OFF-activity, that is $\neg[{\rected[p]}]\equiv\rect[-p]$.
Notice that, by convention, activities are always positive-valued.  Conversely, we refer to
(membrane-) \emph{potential} to characterize a neuron's state.  We observe that the potential
or state of a neuron can be be positive-valued and negative-valued.  
\subsection{Lateral inhibition}
The ''critical race'' described above occurs between excitatory activities originating 
from movement detectors (\eq[advancedphoto]), and laterally propagating inhibitory waves \CITE{RindBramwell96}.
Laterally propagating waves are generated by nearest neighbor coupling such that adjacent
neurons can exchange their membrane potential. In organisms, such coupling is established
by electrical synapses or gap junctions.  Mathematically, it is modeled by
reaction-diffusion systems, which we call diffusion layers.
There exists one diffusion layer ($s \in \{\pos[s],\neg[s]\}$) for each pathway (ON and OFF):
\begin{eqnarray}\label{advancedsyncytium}
	\ddt[s_{ij}]  =  \gLeak (\Vrest-s_\mathit{ij}) & + g_\mathit{exc,ij}(1-s_\mathit{ij})\\ 
		                                       & + D\,\Lap s_\mathit{ij}\nonumber
\end{eqnarray}
where $\gLeak=10$ is related to the diffusion length constant of the system (e.g. \CITE{BendaEtAl01}).
$\Vrest=-0.001$ is a resting potential (the value which is approached by $s_{ij}$ without input).
$D=170$ is the diffusion coefficient, which specifies the speed of the propagation process.
Finally, $\Lap$ implements the Laplacian operator (we used a discrete four point approximation). 
ON- and OFF diffusion layer neurons receive excitatory input $\pos[g]_\mathit{exc}\equiv250\cdot\rect[{\pos[v]}]$
and $\neg[g]_\mathit{exc}\equiv250\cdot\rect[{\neg[v]}]$, respectively (excitatory units $v$ are defined below).
Diffusion layer outputs are given by $\pos[{\rected[s]}]\equiv\rect[{\pos[s]}]$ (ON)
and $\neg[{\rected[s]}]\equiv\rect[{\neg[s]}]$ (OFF), respectively.
\subsection{Summing units}
Summing units are required to evaluate the result of the ''critical race''
between laterally propagating inhibition (provided by input $g_\mathit{inh,ij}$)
and excitatory activity from movement detectors (provided by input $g_\mathit{exc,ij}$).
Each pathway (i.e. ON and OFF) has its own layer of excitatory neurons:
\begin{equation}\label{summingUnits}
	\begin{split}
	\ddt[v_{ij}] =  \gLeak (\Vrest-v_\mathit{ij}) & + g_\mathit{exc,ij}(1-v_\mathit{ij})\\ 
	                                              & -  g_\mathit{inh,ij}(0.25+v_\mathit{ij})
	\end{split}
\end{equation}
where $\gLeak=100$, and $\Vrest=-0.001$.  Excitatory input into summing units is provided by
ON- and OFF-movement detectors, respectively:
\begin{equation}\label{excinput}
	g_\mathit{exc,ij}=250\,{\rected[p]}\cdot\exp(-\xi\,{\rected[s]}_{ij})
\end{equation}
with a gain constant $\xi=500$ (gain constants correspond to synaptic weights). \Eq[excinput]
establishes that diffusion layer activity $\rected[s]_{ij}$ decreases the excitatory input.
In addition, diffusion layer activity drives the potential of the summing units
away from the response threshold  according to
\begin{equation}\label{inhinput}
	g_\mathit{inh,ij}=\xi\,{\rected[s]_{ij}}
\end{equation}
The output of excitatory ON-units is given by $\pos[{\rected[v]}]\equiv\rect[{\pos[v]}]$,
and  $\neg[{\rected[v]}]\equiv\rect[{\neg[v]}]$ (OFF-unit).
Notice that, in contrast to previous modeling attempts \CITE{RindBramwell96,MatsAngel03gc},
in the present approach feedback inhibition is used.  In combination with \eq[excinput],
feedback inhibition ensures that the diffusion layers are not ''drowned in activity'' as a
consequence of their small leakage conductances (c.f. \eq[advancedsyncytium]; notice
that the leakage conductance also specifies diffusion length, and therefore cannot
be chosen arbitrarily high).  The ''drowning effect'' typically
occurs with feedforward circuits, where MD activity directly excites the diffusion layer.
In the latter case, the presence of strong background movement in video sequences
makes the model blind for subsequent object approaches.
\subsection{LGMD neurons} 
Again, each pathway is associated with one LGMD.  The ON-LGMD integrates activity
from excitatory ON-units $\rected[v] \equiv \pos[{\rected[v]}]$, and the OFF-LGMD
neuron integrates activity from OFF-units $\rected[v] \equiv \neg[{\rected[v]}]$,
according to
\formula[excLGMDONOFF1][{ \ge[] = \sum_{i,j}^n\rected[v]_\mathit{ij}}]
The ON- and OFF-LGMD dynamics obeys (notice that $l$ is a scalar variable)
\begin{equation}\label{LGMDONOFF1}
	\ddt[l] = \gLeak (\Vrest-l) +	\gamma_\mathit{ex} \ge[](t) (1-l)
\end{equation}
where $\gLeak=50$, and $\Vrest=-0.001$.  The output of the ON-LGMD
is $\pos[{\rected[l]}](t)\equiv\rect[{\pos[l](t)}]$, and the output
of the OFF-LGMD is  $\neg[{\rected[l]}](t)\equiv\rect[{\neg[l](t)}]$.
$\gamma_\mathit{ex}=5\cdot 128^2/n^2$ is a gain factor, which depends on the size
of the luminance images (we used input images with  $n$ rows and $n$ columns, see below).
\def\mcar{\emph{mcar}}
\def\star{\emph{starwars}}
\def\soft{\emph{softcrash}}
\def\pede{\emph{pedestrian}}
\def\high{\emph{highway}}
\def\res#1{ (resolution #1$\times$#1 pixels) }

\section{Material and methods}
\label{sect:MaterialMethods}
All video sequences were coded with 8 bit resolution per pixel, and had equal numbers of rows and columns.
Some of our video sequences had an unknown frame rate (\emph{mcar} and \emph{starwars}).  The
remaining videos (\emph{softcrash}, \emph{pedestrian}, and \emph{highway}) were recorded with
25 frames per second.
With each video sequence, we recorded activities of the ON-LGMD $\pos[{\rected[l]}](t)$ and
OFF-LGMD $\pos[{\rected[l]}](t)$.  In what follows, we give a brief description of each video
(see also \fig[videos]):
\begin{description}
\item[mcar]
The \mcar\s video \res{285} shows an approaching car to a still observer.
Since the observer does not move, no background movement is generated
in this sequence.
\item[starwars]
The \star\s video \res{285} shows three approaching space ships.  The 
background is highly textured and moves in opposite direction to 
the space ships.  The approaching space ships contrast only weakly 
with the background.  This video also shows interlacing artifacts,
and high frequency noise.
\item[softcrash]
The \soft\s video \res{150} was recorded from a driving car hitting
a non-rigid obstacle.
\item[pedestrian]
The \pede\s video \res{150} was recorded from a slowly driving
car, where a pedestrian suddenly appears and runs across the scene.
\item[highway]
The \high\s video \res{150} video was recorded from a car driving across a city
highway.  This video contains track changes, and other cars driving on adjacent lanes
or ahead, respectively.
\end{description}
\begin{figure*}[ht!]
	\begin{minipage}[b]{\textwidth}
		\begin{center}\scalebox{0.85}{\includegraphics{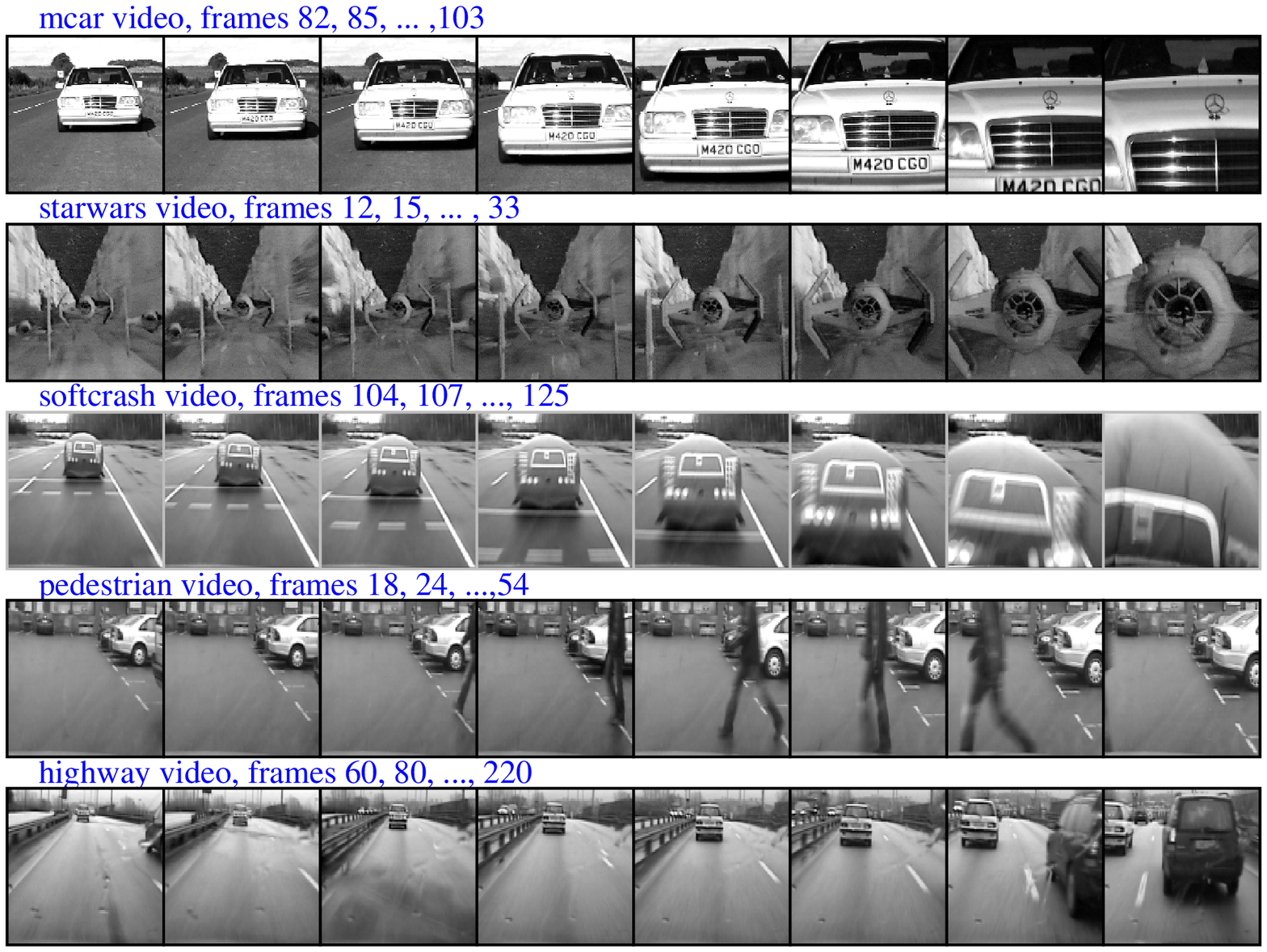}}\end{center}
	\end{minipage}
	\Caption[videos][Video sequences][{The figure shows single frames from the video sequences which were used for the simulations.}]
\end{figure*}
\section{Results}
\label{sect:results}
\begin{figure}[h!]
	\centering
	\includegraphics[width=2.5in]{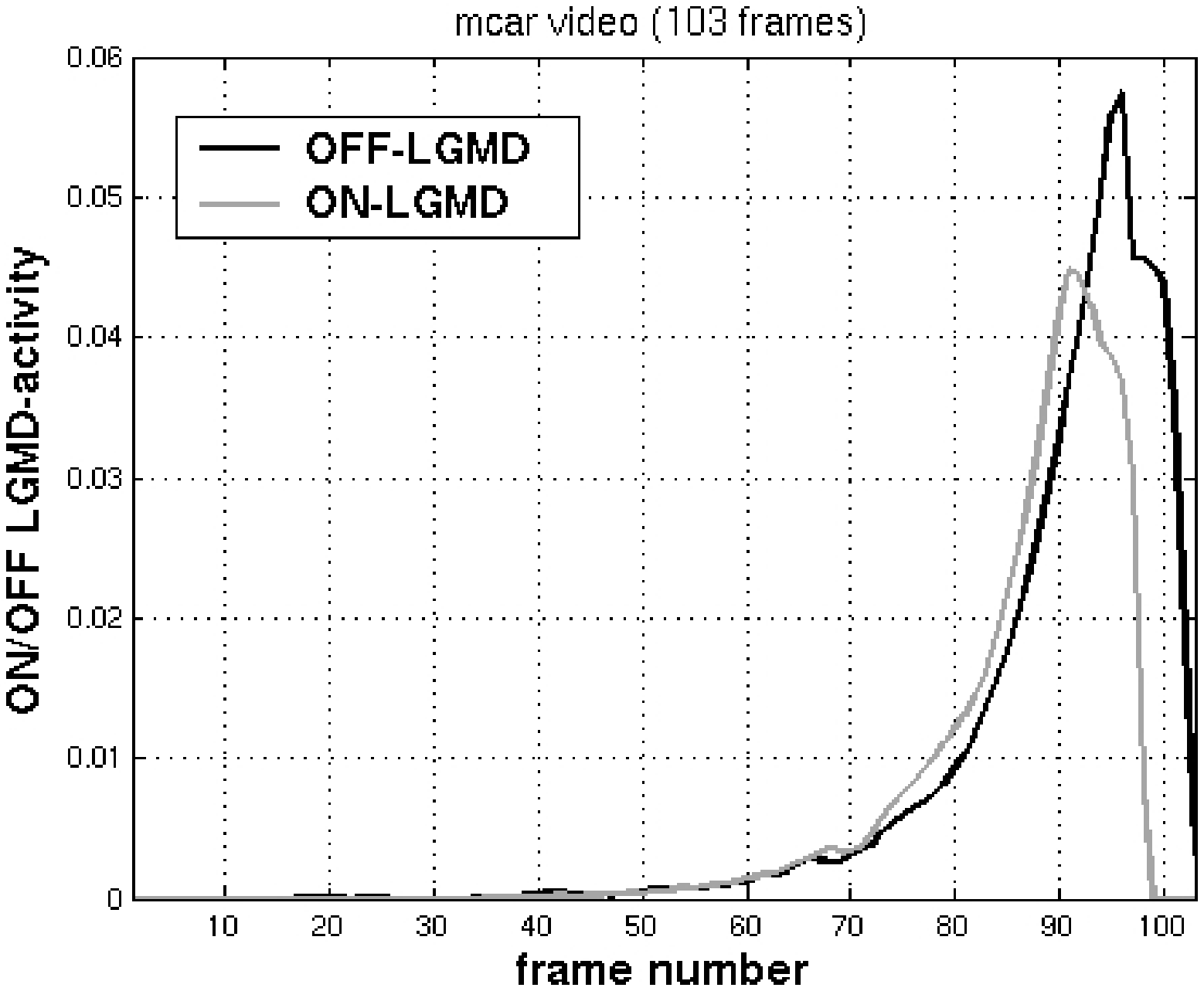}
	\vspace*{-0.5cm}\Caption[mcar][Results for the \mcar\s video][{}]
\end{figure} 
Figures \ref{mcar} to \ref{highway} show the output of the model for the various video
sequences shown in \fig[videos].  The \mcar\s video represents the easiest test for
the model, since there is virtually no background movement involved.  Activities
of both LGMDs (\fig[mcar]) smoothly follow the approaching car, where the curve
for the ON-LGMD reaches a maximum before the OFF-LGMD.  Triggering a collision
alert in this case is rather easy, since one only needs to detect the rising phase
of LGMD responses.  Notice that both curves strongly resemble the \etafun\s (see
introduction).\\
\begin{figure}[h!]
	\centering
	\includegraphics[width=2.5in]{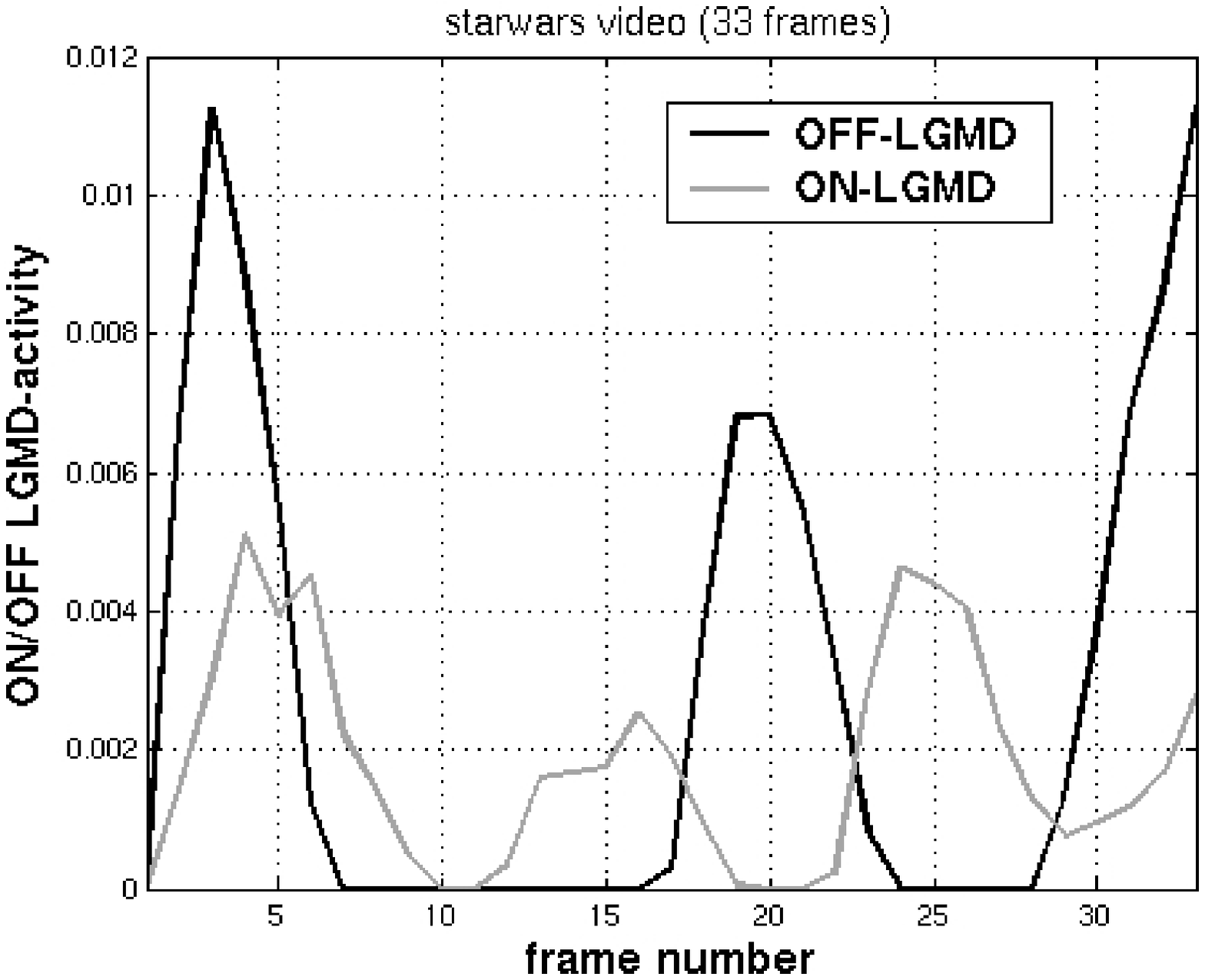}
	\vspace*{-0.5cm}\Caption[starwars][Results for the \star\s video][{}]
\end{figure} 
Triggering a collision alert seems more difficult when monitoring LGMD activities
with the \star\s video (\fig[starwars]).  At the beginning, a strong OFF-response, and a weaker
ON-response are seen.  These are transient effects, created by strong background movement.
Since inhibitory activity has not built up and spreaded laterally yet,
background movement is not suppressed.  The OFF-peak in the middle results from
the approaching left and right space ship accompanying the middle one, as they
move out of the scene.  In the last frames, the OFF-response increases
more strongly compared to the ON-response as the center space ship
grows larger than the frame. The last ON-peak
denotes the period where the lateral space ships have just moved out
of the scene (beginning) until the center space ship fills the whole
frame (end).
Hence, with the \star\s video, collision detection is complicated because of
the presence of secondary peaks. Both ON- and OFF-response amplitudes are
substantially diminished by feedback inhibition.\\
\begin{figure}[h!]
	\centering
	\includegraphics[width=2.5in]{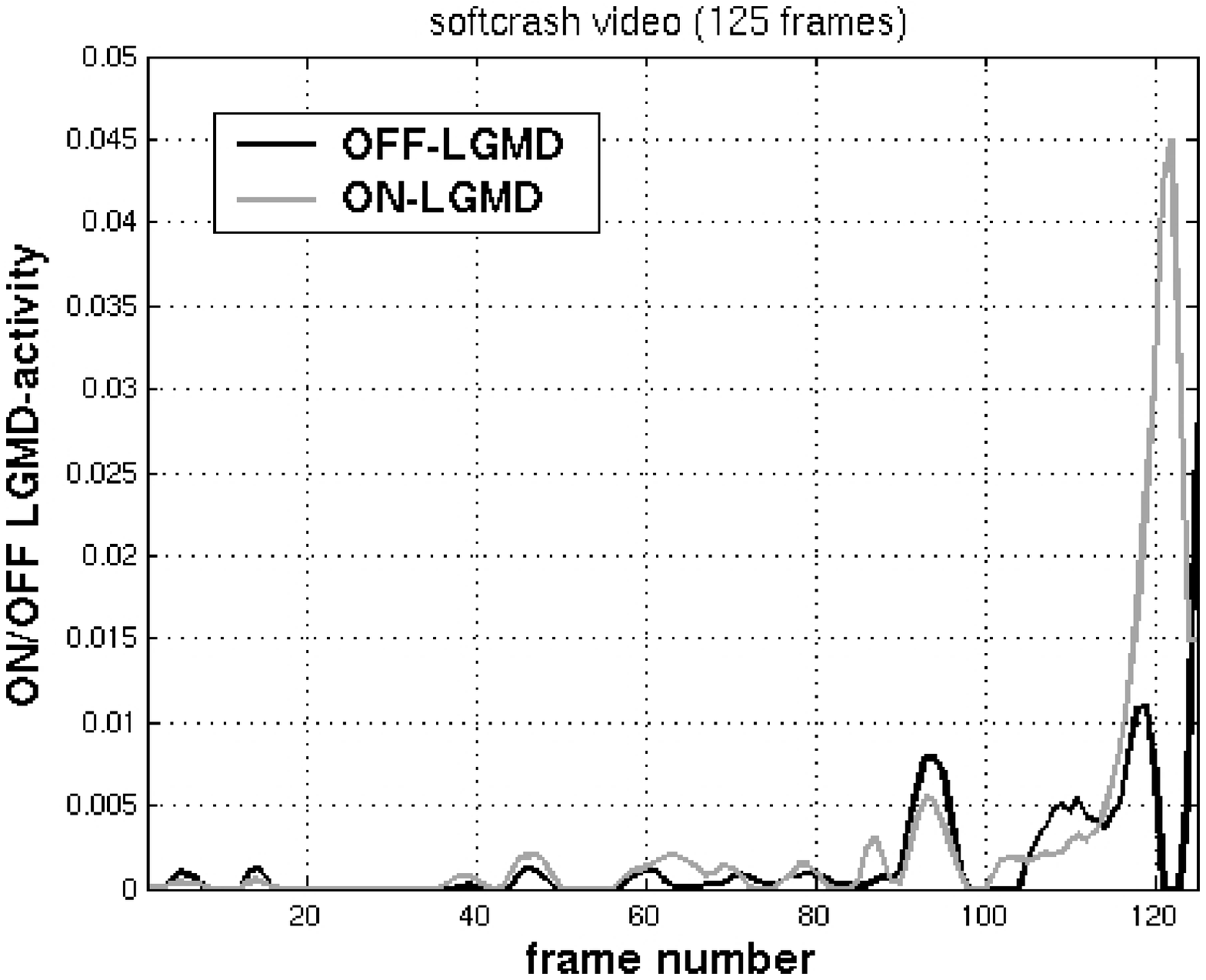}
	\vspace*{-0.5cm}\Caption[softcrash][Results for the \soft\s video][{}]
\end{figure} 
The \soft\s video (\fig[softcrash]) shows several secondary response peaks
in LGMD activities. These peaks reflect novel background activity (rising
phase), which is caught up by lateral inhibition immediately (falling phase). 
The amplitudes of these peaks are relatively small and increase towards the
collision event.  The ON-LGMD response shows a pronounced maximum at collision
time, what makes collision detection feasible in this situation.\\
\begin{figure}[h!]
	\centering
	\includegraphics[width=2.5in]{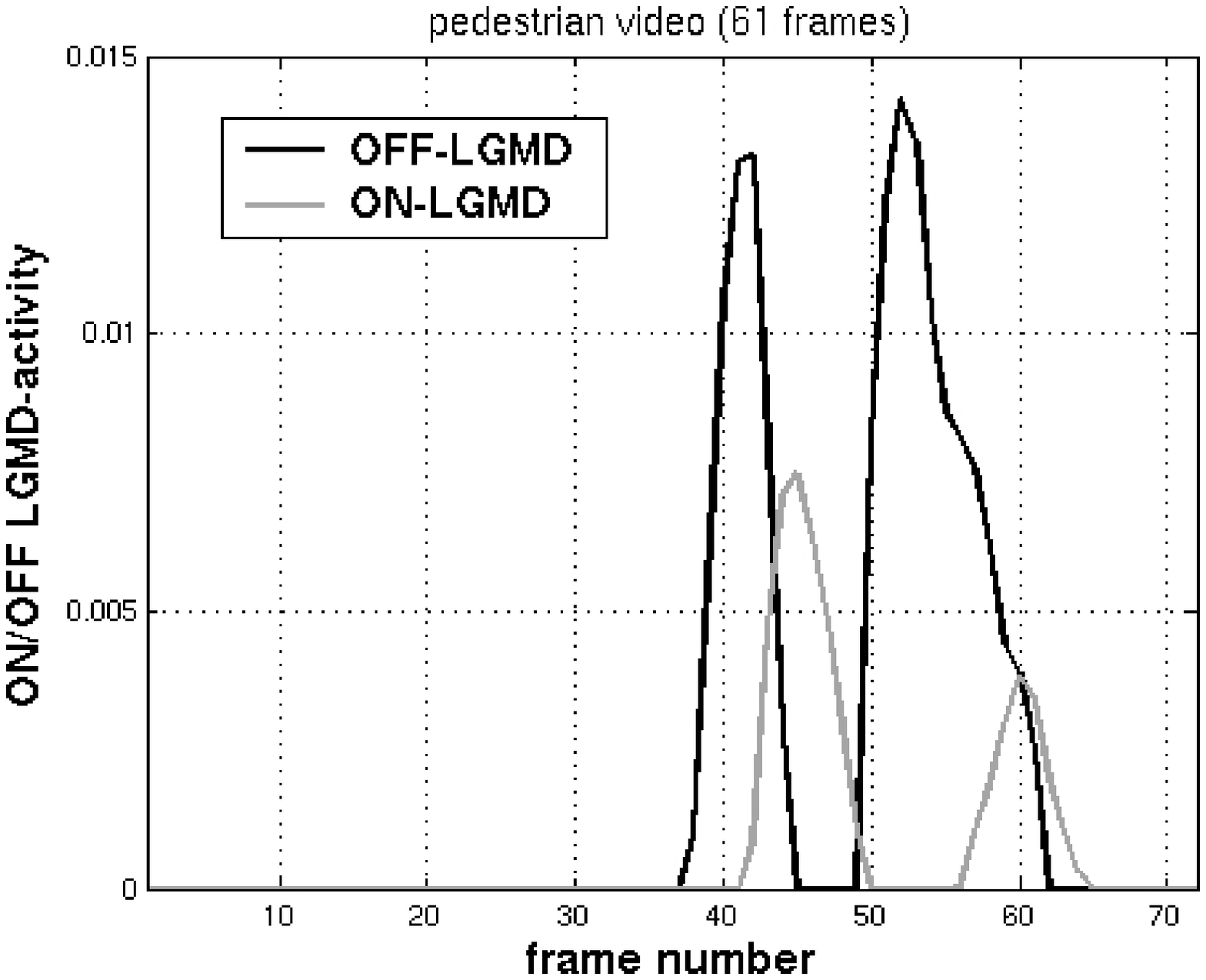}
	\vspace*{-0.5cm}\Caption[pedestrian][Results for the \pede\s video][{}]
\end{figure} 
The model may also serve to detect a pedestrian crossing a scene (\fig[pedestrian], \pede\s video).
The model shows distinct response peaks when the pedestrian enters the scene.
However, lateral inhibition soon suppresses corresponding responses.  Since
lateral inhibition cuts its own excitation by means of \eq[excinput], response
suppression occurs until inhibition dissipated, and subsequently LGMD responses
may form again (second peak).  Notice that since the car drives relatively slow,
LGMD responses due to background movement can be neglected.\\
\begin{figure}[h!]
	\centering
	\includegraphics[width=2.5in]{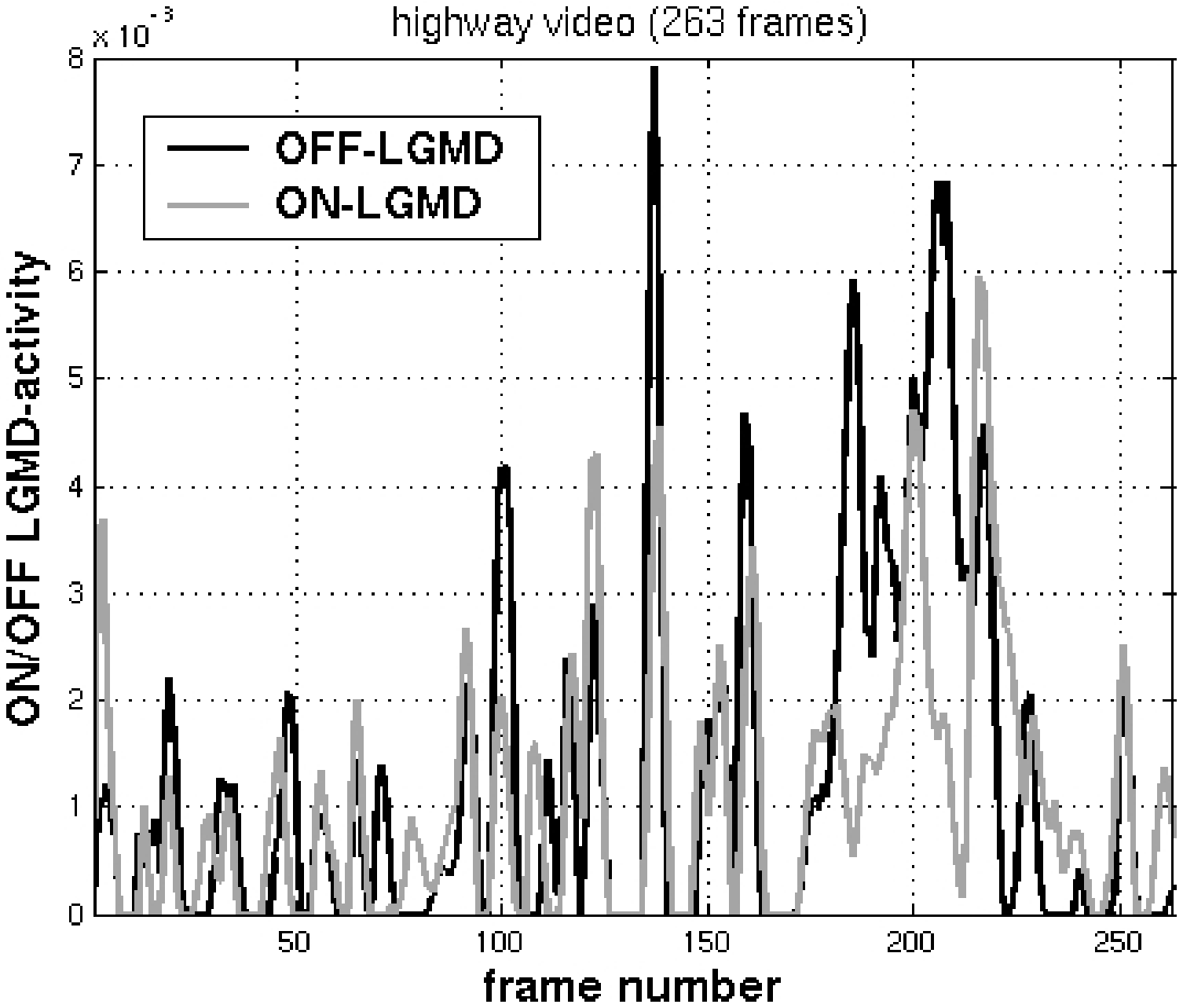}
	\vspace*{-0.5cm}\Caption[highway][Results for the \high\s video][{}]
\end{figure} 
Ideally there would be no model response with the \high\s video (\fig[highway]),
because of the absence of any collision situation.  All LGMD responses are due
to background movement (such as street signs), cars passing by etc.  This is
reflected in the response amplitudes, which are comparatively small (c.f. \fig[starwars]),
and are attenuated by lateral inhibition.
\section{Discussion and conclusions}
\label{discussion}
In this paper we presented an approach to collision detection, based on know neurophysiological
data of the \emph{lobula giant movement detector} (LGMD).  The present approach involves several
improvements with respect to a previously presented model \CITE{MatsAngel03gc}. In the previous
model, lateral inhibition was of feedforward type, that is movement detectors directly supported
inhibition.  The problem with the latter circuit is that strong background movement soon floods
the diffusion layer associated with lateral inhibition.  As a consequence, LGMD responses cease, and the model gets
blind against object approaches.  To remedy this problem, two parallel channels (ON and OFF) where
introduced, with the goal to reduce the overall activity in diffusion layers.  Furthermore, 
lateral inhibition within the present approach is of the feedback type, with the additional effect
that once inhibition is active, it is cut off from is own excitation, such that it cannot grow further, but
rather dissipates.  In this way flooding of diffusion layers due to strong background movement
is effectively reduced.  Despite of it all, the present approach as it stands lacks several
additional mechanisms.\\
First, the present model also responds to self-motion of agents.  Such responses could possibly be attenuated
by including feedforward inhibition, as described in the introduction.  For automotive applications,
it will be sufficient to include a detector for left and right movement (either visually by means
of directional movement detectors, or by an external signal, e.g. from the steering wheel).\\
Second, a decision mechanism has to evaluate ON-LGMD and OFF-LGMD responses in order to detect
a collision event or not.  The results shown indicate that this is a non-trivial problem, since
other situations than collisions may also trigger large LGMD responses (e.g. any sufficiently big
object which approaches, but finally does not collide).\\
Third, in order to make the model independent from object contrasts or illumination, adaptational
mechanisms at the movement detector (or photoreceptor) level have to be included.  Simple saturation
of photoreceptor responses may only be viable strategy if noise levels associated with luminance
changes are negligible.\\
\section*{Acknowledgements}
This work has been supported by the LOCUST project (IST2001-38097) and the VISTA
project (TIC2003-09817-C02-01).
The authors like to thank F.C.Rind for kindly providing the video sequences
\emph{mcar} and \emph{starwars}, and Volvo Car Corporation for the video sequences
\emph{softcrash}, \emph{pedestrian}, and \emph{highway}.
\bibliography{../../refs}   
\bibliographystyle{unsrt}
\end{document}

%% file: definitions.tex
\def\Caption[#1][#2][#3]{\caption{\label{#1}\small{{\bf #2.} #3}}}

\newcommand{\cA}{{\cal A}}
\newcommand{\cB}{{\cal B}}
\newcommand{\cC}{{\cal C}}
\newcommand{\cD}{{\cal D}}
\newcommand{\cE}{{\cal E}}
\newcommand{\cF}{{\cal F}}
\newcommand{\cG}{{\cal G}}
\newcommand{\cH}{{\cal H}}
\newcommand{\cI}{{\cal I}}
\newcommand{\cJ}{{\cal J}}
\newcommand{\cK}{{\cal K}}
\newcommand{\cL}{{\cal L}}
\newcommand{\cM}{{\cal M}}
\newcommand{\cN}{{\cal N}}
\newcommand{\cO}{{\cal O}}
\newcommand{\cP}{{\cal P}}
\newcommand{\cQ}{{\cal Q}}
\newcommand{\cR}{{\cal R}}
\newcommand{\cS}{{\cal S}}
\newcommand{\cT}{{\cal T}}
\newcommand{\cU}{{\cal U}}
\newcommand{\cV}{{\cal V}}
\newcommand{\cW}{{\cal W}}
\newcommand{\cX}{{\cal X}}
\newcommand{\cY}{{\cal Y}}
\newcommand{\cZ}{{\cal Z}}
\newcommand{\dd}{\,d}

\newcommand{\Res}{\mathit{Res}}
\newcommand{\where}{\ \ \ \mathit{where}\ }
\newcommand{\with}{\ \ \ \mathit{with}\ }
%
\def\ddt[#1]{\frac{d#1(t)}{dt}}
\def\partialddt[#1]{\frac{\partial#1}{\partial t}}
%
\def\formula[#1][#2]{\begin{equation}\label{#1}#2\end{equation}}
%
%
\def\SYX[#1][#2]{$#1_\mathit{#2}$}
\def\syx[#1][#2]{#1_\mathit{#2}}
%
\def\tab[#1]{table~\ref{#1}}
\def\Tab[#1]{Table~\ref{#1}}
\def\cftab[#1]{(cf. table~\ref{#1})}
%
\def\eq[#1]{equation~\ref{#1}}
\def\Eq[#1]{Equation~\ref{#1}}
\def\eqPair[#1][#2]{equations~\ref{#1} and \ref{#2}}
\def\EqPair[#1][#2]{Equations~\ref{#1} and \ref{#2}}
\def\eqs[#1][#2]{equation~\ref{#1} to \ref{#2}}
\def\Eqs[#1][#2]{Equation~\ref{#1} to \ref{#2}}
\def\cfeq[#1]{(cf. equation~\ref{#1})}
%
\def\fig[#1]{figure~\ref{#1}}
\def\Fig[#1]{Figure~\ref{#1}}
\def\cffig[#1]{(cf. figure~\ref{#1})}
%
\def\apdx[#1]{appendix~\ref{#1}}
\def\Apdx[#1]{Appendix~\ref{#1}}
\def\cfapdx[#1]{(cf. appendix~\ref{#1})}
%
%
\def\sec[#1]{section~\ref{#1}}
\def\Sec[#1]{Section~\ref{#1}}
\def\cfsec[#1]{(cf. section~\ref{#1})}
%
%
%
\def\chap[#1]{chapter~\ref{#1}}
\def\Chap[#1]{Chapter~\ref{#1}}
\def\cfchap[#1]{(cf. chapter~\ref{#1})}
%
%
\def\s{{ }}
\def\bs{$\!\!$}
%
%
\def\img[#1]{{\scriptsize {\bf #1}}}
\def\size[#1]{(size #1$\times$#1 pixels)}
\def\Picard{{\scriptsize {\bf Picard}} }
\def\duBuf{{\scriptsize {\bf du Buf}} }
\def\Lena{{\scriptsize {\bf Lena}} }
\def\Bart{{\scriptsize {\bf Bart}} }
\def\Homer{{\scriptsize {\bf Homer}} }
\def\Panther{{\scriptsize {\bf Panther}} }
\def\Peppers{{\scriptsize {\bf Peppers}} }
\def\Pyramid{{\scriptsize {\bf Pyramid}} }
\def\Ehrenstein{{\scriptsize {\bf Ehrenstein}} }
\def\Boats{{\scriptsize {\bf Boats}} }
\def\MIT{{\scriptsize {\bf MIT}} }
\def\Le{{\bf Left: }} 
\def\Mi{{\bf Middle: }} 
\def\Ri{{\bf Right: }} 
\def\le{{\bf left: }} 
\def\mi{{\bf middle: }} 
\def\ri{{\bf right: }} 
\def\permission[#1]{(Reprinted {\it without} any permission from \cite{#1})}
%
\def\Ba{{\bf (a)} }
\def\Bb{{\bf (b)} }
\def\Bc{{\bf (c)} }
\def\Bd{{\bf (d)} }
\def\Be{{\bf (e)} }
\def\Bf{{\bf (f)} }
\def\Bg{{\bf (g)} }
\def\Bh{{\bf (h)} }
%
\def\Ia{{\it a} }
\def\Ib{{\it b} }
\def\Ic{{\it c} }
\def\Id{{\it d} }
\def\Ie{{\it e} }
\def\If{{\it f} }
\def\Ig{{\it g} }
\def\Ih{{\it h} }
%
%
%
%
%
\def\sm[#1]{{\small $#1$}}
\def\KochbuchRef[#1]{(see \cite{Kochbuch99}, chapter {\it #1} with references)}
\def\Kochbuch[#1]{(\cite{Kochbuch99}, chapter {\it #1})}
\def\MelRef[#1]{(see \cite{MelReview94}, section {\it #1} with references)}
\def\Mel[#1]{(\cite{MelReview94}, section {\it #1})}
\def\KapLeeShap[#1]{(\cite{KaplanEtAl90}, section {\it #1})}
\def\Webvision[#1][#2]{\cite{KolbEtAl#1}, section {\it #2}}
\def\RecipesC[#1]{(\cite{Recipes}, section {\it #1})}
\def\RecipesP[#1]{(\cite{Recipes}, page #1)}
%
%
\def\taueff{\tau_\mathit{eff}}
\def\gEff{g_\mathit{eff}}
\def\V[#1]{V\!_\mathit{#1}}
\def\Vth{V_\mathit{th}}
\def\Vinf{V_{\infty}}
\def\Vinfm{V_{m,\infty}}
\def\Vrest{V_\mathit{rest}}
\def\gLeak{g_\mathit{leak}}
\def\gSyn{g_\mathit{syn}}
\def\gSyni{g_\mathit{syn,i}}
\def\gSynx[#1]{g_\mathit{syn,#1}}
\def\Esyn{E_\mathit{syn}}
\def\Esyni{E_\mathit{syn,i}}
\def\Esynx[#1]{E_\mathit{syn,#1}}
\def\Isyn{I_\mathit{syn}}
%
%
\def\Egaba[#1]{E_\mathit{GABA_{#1}}}
\def\gGaba[#1]{g_\mathit{GABA_{#1}}}
\def\Enmda{E_\mathit{NMDA}}
\def\gNmda{g_\mathit{NMDA}}
\def\Eampa{E_\mathit{AMPA}}
\def\gAmpa{g_\mathit{AMPA}}
\def\GABA[#1]{GABA$_\mathit{#1}$}
%
%
\def\Mg{{\mathit Mg}^{2+}}
\def\Ca{{\mathit Ca}^{2+}}
\def\Na{{\mathit Na}^+}
\def\Cl{{\mathit Cl}^-}
\def\K{{\mathit K}^+}
\def\A{{\mathit A}^-}
%
%
\def\Eos{E_\mathit{os}}
\def\Ee{E_\mathit{ex}}
\def\Ei{E_\mathit{in}}
\def\Esi{E_\mathit{si}}
\def\ge[#1]{\mathcal{E}_\mathit{#1}}
\def\gi[#1]{\mathcal{I}_\mathit{#1}}
\def\gsi[#1]{\gi[#1,si]}
\def\gos[#1]{g_\mathrm{#1}\mathit{,os}}
%
%
\def\sig[#1]{\sigma\!_\mathrm{#1}}
\def\Gauss[#1]{\mathcal{G}_\sig[#1]}
\def\Gaussij[#1][#2]{\Gauss[#1] \mathit{\!(#2)}}
\def\GaussII{\mathcal{G}_{\sig[h]\sig[v]}}
%
%
\def\conv{\otimes}
%
%
\def\rect[#1]{\left[#1\right]^+}	
\def\rected[#1]{\tilde{#1}}		
%
%
\def\Del{\mathbf{\nabla}}
\def\Lap{\nabla^2}
%
%
\def\heaviside{H}
%
%
\def\normalize[#1]{\mathit{Norm}\!\left[ #1 \right]}
\def\neighborhood[#1]{{\mathcal N}\!_\mathit{#1}}
%
%
\newcommand{\masy}{max-syncytium }
\newcommand{\misy}{min-syncytium }
\newcommand{\Masy}{Max-syncytium }
\newcommand{\Misy}{Min-syncytium }
\newcommand{\masya}{max-syncytia }
\newcommand{\misya}{min-syncytia }
\newcommand{\Masya}{Max-syncytia }
\newcommand{\Misya}{Min-syncytia }
\def\KEIL{\mathcal K}
\def\KEILIAN[#1]{\KEIL_{\!#1}}
\def\keilian{\KEILIAN[\lambda]}
\def\poskeil{\KEIL^+}
\def\negkeil{\KEIL^-}
%
%
\def\perm{D}
\def\permvar{D(\BOUNDARY[])}
\def\permvarGrossberg[#1][#2]{\perm/(1+\epsilon(\BOUNDARY[#1]+\BOUNDARY[#2]))}
\def\discretePermDiff[#1][#2][#3]{\sum_{\mathit{(#1,#2)}\in\neighborhood[#3]}
(x_\mathit{#1#2}-x_\mathit{#3}) \perm\!_{#1 #2 #3}} 
%
%
\def\boundarySymbol{w}
\def\boundary[#1]{\boundarySymbol\!_\mathit{#1}}
\def\BOUNDARY[#1]{\rected[\boundarySymbol]_\mathit{#1}}
%
%
%
%
\def\Lumi[#1]{\mathcal{L}_\mathit{#1}}
\def\perceivedLumi[#1]{\hat{\mathcal{L}}_\mathit{#1}}
\def\Bright[#1]{\mathcal{B}_\mathit{#1}}
\def\Contrast[#1]{\mathcal{C}_\mathit{#1}}
\def\L2C[#1]{M_\mathit{#1}}			
\def\iL2C[#1]{\L2C[#1]^{-1}}
%
%
\def\pos[#1]{#1^{\,\oplus}}		
\def\neg[#1]{#1^{\,\ominus}}		
%
%
\def\xcell{x}	
\def\ycell{y}	
\def\mtplx{m}	

\def\XCELL{\rected[\xcell]}	
\def\YCELL{\rected[\ycell]}	
\def\MTPLX{\rected[\mtplx]}	

\def\xon[#1]{\pos[\xcell\!_\mathit{#1}]}
\def\xoff[#1]{\neg[\xcell\!_\mathit{#1}]}
\def\XON[#1]{\pos[\XCELL\!_\mathit{#1}]}
\def\XOFF[#1]{\neg[\XCELL\!_\mathit{#1}]}

\def\yon[#1]{\pos[\ycell\!_\mathit{#1}]}
\def\yoff[#1]{\neg[\ycell\!_\mathit{#1}]}
\def\YON[#1]{\pos[\YCELL\!_\mathit{#1}]}
\def\YOFF[#1]{\neg[\YCELL\!_\mathit{#1}]}

\def\mon[#1]{\pos[\mtplx\!_\mathit{#1}]}
\def\moff[#1]{\neg[\mtplx\!_\mathit{#1}]}
\def\MON[#1]{\pos[\MTPLX\!_\mathit{#1}]}
\def\MOFF[#1]{\neg[\MTPLX\!_\mathit{#1}]}
%
%
\def\Xon{ON$_\mathrm{x}$}
\def\Xoff{OFF$_\mathrm{x}$}
\def\Yon{ON$_\mathrm{y}$}
\def\Yoff{OFF$_\mathrm{y}$}
\def\yy{ON$_\mathrm{y}\times$OFF$_\mathrm{y}$ }
%
%
\def\retina{r}			
\def\RETINA{\rected[\retina]}	

\def\on[#1]{\pos[\retina\!_\mathit{#1}]}
\def\off[#1]{\neg[\retina\!_\mathit{#1}]}
\def\ON[#1]{\pos[\RETINA_\mathit{#1}]}
\def\OFF[#1]{\neg[\RETINA_\mathit{#1}]}
%
%
\def\lcc{\Lambda}
\def\lccON[#1]{\ON[{step\# #1}]}
\def\lccOFF[#1]{\OFF[{step\# #1}]}
%
%
\def\gLeakON{\pos[\gLeak]}
\def\gLeakOFF{\neg[\gLeak]}
\def\gLeakONI{\pos[\tilde{g}_\mathit{leak}]}
\def\gLeakOFFI{\neg[\tilde{g}_\mathit{leak}]}

\def\gsiON[#1]{\pos[g_{#1,si}]}
\def\gsiOFF[#1]{\neg[g_{#1,si}]}

\def\gosON[#1]{\pos[g](\os[#1])}
\def\gosOFF[#1]{\neg[g](\os[#1])}
%
%
\def\cent[#1]{\mathcal{C}\!_\mathit{#1}}
\def\surr[#1]{\mathcal{S}\!_\mathit{#1}}
\def\os[#1]{\mathcal{O\!S}\!_\mathit{#1}}
\def\indxCent{_\mathrm{c}}
\def\indxSurr{_\mathrm{s}}
\def\indxOs{_\mathrm{os}}
\def\onDOG[#1]{\rect[{\cent[#1]-\surr[#1]}]}
\def\offDOG[#1]{\rect[{\surr[#1]-\cent[#1]}]}
\def\DOG{\mathcal{D}\!\mathrm{o}\mathcal{G}}

\def\gain{\gamma}
\def\mgainON[#1]{\pos[\gain](\os[#1])}
\def\mgainOFF[#1]{\neg[\gain](\os[#1])}
\def\offsetON{\pos[\gain\mathit{0}]}
\def\offsetOFF{\neg[\gain\mathit{0}]}

\def\mExpSigConst{\lambda_\mathrm{(3)}}	
\def\mExpSigConstON{\pos[\mExpSigConst]}
\def\mExpSigConstOFF{\neg[\mExpSigConst]}

\def\mTanhConst{\lambda_\mathrm{(4)}}	
\def\mTanhConstON{\pos[\mTanhConst]}
\def\mTanhConstOFF{\neg[\mTanhConst]}

\def\mExpConst{\lambda_\mathrm{(5)}}	
%
%
\def\con{{\it c{\bf on}tour}}
\def\coff{{\it c{\bf off}tour}}

\def\cons{{\it c{\bf on}tours}}
\def\coffs{{\it c{\bf off}tours}}

\def\CON{{\it C{\bf on}tour}}
\def\COFF{{\it C{\bf off}tour}}

\def\CONs{{\it C{\bf on}tours}}
\def\COFFs{{\it C{\bf off}tours}}
%
%
\def\o{o}
\def\O{\rected[o]}
\def\oon[#1]{\pos[\o_\mathrm{#1}]}
\def\ooff[#1]{\neg[\o_\mathrm{#1}]}
\def\OON[#1]{\pos[\O_\mathrm{#1}]}
\def\OOFF[#1]{\neg[\O_\mathrm{#1}]}
%
%
\def\ld{ld}
\def\dl{dl}
\def\ldl{ldl}
\def\dld{dld}

\def\LD{\mathrm{\ld}}
\def\DL{\mathrm{\dl}}
\def\LDL{\mathrm{\ldl}}
\def\DLD{\mathrm{\dld}}
%
%
\def\Lon[#1]{\pos[\mathcal{L}_\mathrm{#1}]}
\def\Loff[#1]{\neg[\mathcal{L}_\mathrm{#1}]}
\def\Ron[#1]{\pos[\mathcal{R}_\mathrm{#1}]}
\def\Roff[#1]{\neg[\mathcal{R}_\mathrm{#1}]}
\def\Con[#1]{\pos[\mathcal{C}_\mathrm{#1}]}
\def\Coff[#1]{\neg[\mathcal{C}_\mathrm{#1}]}
\def\satCon[#1]{\pos[C_\mathrm{#1}]}
\def\satCoff[#1]{\neg[C_\mathrm{#1}]}
%
%
\def\brightsym{\circ}
\def\darksym{\bullet}
\def\bright[#1][#2]{#1^\mathrm{#2\brightsym}}	
\def\dark[#1][#2]{#1^\mathrm{#2\darksym}}		
%
%
\def\texture{t}
\def\textout{T}
\def\TEXTURE{\rected[\texture]}
\def\ton[#1][#2]{\bright[\texture_\mathrm{#1}][#2]}	
\def\toff[#1][#2]{\dark[\texture_\mathrm{#1}][#2]}
\def\TON[#1][#2]{\bright[\TEXTURE_\mathrm{#1}][#2]}
\def\TOFF[#1][#2]{\dark[\TEXTURE_\mathrm{#1}][#2]}
\def\outTON[#1]{\bright[\textout_\mathrm{#1}][]}
\def\outTOFF[#1]{\dark[\textout_\mathrm{#1}][]}
%
%
\def\gradient{g}
\def\gradout{G}
\def\GRADIENT{\rected[\gradient]}
\def\grad[#1][#2]{\gradient_\mathrm{#1}^\mathrm{(#2)}} 	
\def\GRAD[#1][#2]{\GRADIENT\mathrm{#1}^\mathrm{((#2)}}	

\def\gon[#1][#2]{\bright[\gradient_\mathrm{#1}][#2]}	
\def\goff[#1][#2]{\dark[\gradient_\mathrm{#1}][#2]}
\def\GON[#1][#2]{\bright[\GRADIENT_\mathrm{#1}][#2]}
\def\GOFF[#1][#2]{\dark[\GRADIENT_\mathrm{#1}][#2]}
\def\outGON[#1]{\bright[\gradout_\mathrm{#1}][]}
\def\outGOFF[#1]{\dark[\gradout_\mathrm{#1}][]}
\def\featureZONE[#1]{\mathcal{F}_\mathrm{#1}}		
\def\gradientZONE[#1]{\mathcal{G}_\mathrm{#1}}		
\def\gammaON[#1]{\bright[\gamma_\mathrm{#1}][]}	
 \def\gammaOFF[#1]{\dark[\gamma_\mathrm{#1}][]}
%
%
\def\hotspot{h}
\def\hotout{H}
\def\hon[#1][#2]{\bright[\hotspot_\mathrm{#1}][#2]}	
\def\hoff[#1][#2]{\dark[\hotspot_\mathrm{#1}][#2]}
\def\outHON[#1]{\bright[\hotout_\mathrm{#1}][]}
\def\outHOFF[#1]{\dark[\hotout_\mathrm{#1}][]}

\def\surface{s}
\def\surfout{S}
\def\SURFACE{\rected[\surface]}
\def\son[#1][#2]{\bright[\surface_\mathrm{#1}][#2]}	
\def\soff[#1][#2]{\dark[\surface_\mathrm{#1}][#2]}
\def\SON[#1][#2]{\bright[\SURFACE_\mathrm{#1}][#2]}
\def\SOFF[#1][#2]{\dark[\SURFACE_\mathrm{#1}][#2]}
\def\outSON[#1]{\bright[\surfout_\mathrm{#1}][]}
\def\outSOFF[#1]{\dark[\surfout_\mathrm{#1}][]}

\def\Spread[#1]{\mathbf{S}_\mathrm{#1}}			

\def\PENALTY{q}
\def\pon[#1][#2]{\bright[\PENALTY_\mathrm{#1}][#2]}	
\def\poff[#1][#2]{\dark[\PENALTY_\mathrm{#1}][#2]}

\def\brightKEIL[#1]{\KEIL_{\!#1}^\brightsym}
\def\darkKEIL[#1]{\KEIL_{\!#1}^\darksym}
\def\ZONE[#1]{\mathcal{Z}_\mathit{#1}}
\def\zone[#1]{z_\mathit{#1}}
\def\thresh[#1]{\mathit{thresh}_\mathrm{#1}}
\def\zoneThresholdValue{\Theta_z}
\def\concoffThresholdValue{\Theta_\boundarySymbol}

\def\contour[#1]{\bright[\boundary[#1]][]}
\def\cofftour[#1]{\dark[\boundary[#1]][]}
\def\concoffgain{\gamma_\boundarySymbol}

\def\PERCEPT{p}
\def\percept[#1]{\PERCEPT_\mathrm{#1}}

%
%
\def\island{\mathcal{I}}
\def\path{\mathcal{P}}
\newtheorem{DEF}{Definition}
%
%
\def\fillIn{f}
\def\FILLIN{\rect[\fillIn]}
\def\brightFillInSteve[#1]{\pos[\fillIn\!_\mathit{#1}]}
\def\darkFillInSteve[#1]{\neg[\fillIn\!_\mathit{#1}]}
\def\BRIGHTFillInSteve[#1]{\pos[\FILLIN_\mathit{#1}]}
\def\DARKFillInSteve[#1]{\neg[\FILLIN_\mathit{#1}]}
%
%
\def\ANDNOT{{\scriptsize {\bf AND-NOT}}}
\def\AND{{\scriptsize {\bf AND}}}
\def\XOR{{\scriptsize {\bf XOR}}}
\def\OR{{\scriptsize {\bf OR}}}

%
%
%
%
\def\dec[#1]{f[\mathit{#1}]}
\def\inc[#1]{g[\mathit{#1}]}
\def\brightFun[#1]{f_\mathit{#1}^\brightsym}	\def\darkFun[#1]{f_\mathit{#1}^\darksym}
\def\brightP[#1]{P_\mathit{#1}^\brightsym}	\def\darkP[#1]{P_\mathit{#1}^\darksym}
%
%
%
%
%
%

%
\def\I{{\it (i)} }
\def\II{{\it (ii)} }
\def\III{{\it (iii)} }
\def\IV{{\it (iv)} }
\def\V{{\it (v)} }
\def\VI{{\it (vi)} }
\def\VII{{\it (vii)} }
\def\VIII{{\it (viii)} }
\def\IX{{\it (ix)} }
\def\X{{\it (x)} }